\documentclass[11pt]{article}
\usepackage{moriond,epsfig}

\def\Journal#1#2#3#4{{#1} {\bf #2}, #3 (#4)}

\def\NIM{\em Nucl. Instrum. Methods}

\def\PLB{{\em Phys. Lett.}  B}
\def\PRL{\em Phys. Rev. Lett.}
\def\PRD{{\em Phys. Rev.} D}

\def\bbb{\bar{B}/B}
\def\la{\Lambda}
\def\al{\bar{\Lambda}}
\def\xim{\Xi^{-}}
\def\axp{\bar{\Xi}^{+}}
\def\omm{\Omega^{-}}
\def\aop{\bar{\Omega}^{+}}
\def\pt{p_{T}}
\def\be{\begin{equation}}
\def\ee{\end{equation}}
\def\bea{\begin{eqnarray}}
\def\eea{\end{eqnarray}}

\begin{document}
\vspace*{4cm}
\title{STRANGE CONTENT OF BARYONS AT RHIC}

\author{ B. HIPPOLYTE for the STAR Collaboration\footnote{for
the full author list, see {\sf http://www.star.bnl.gov/STAR/smd/collab/sci-apr03.ps}.}}
\address{Physics Department, Yale University,\\
P.O. Box 208124, New Haven CT, 06520, USA.}

\maketitle\abstracts{
Via the study of strange particle production within the STAR
experiment, we try to address the surprising amount of baryon transport
at the Relativistic Heavy Ion Collider (RHIC). We report here
preliminary results showing that, at mid-rapidity and for the
top energy of RHIC, the number of created baryons exceeds
the number transported from the colliding nuclei.   
However, thanks to the large acceptance of the experimental
setup, one could expect to observe the transition between the
``soft'' regime (low transverse momentum -$\pt$- region 
corresponding to a bulk of hot and dense matter hadronizing) and
the perturbative one (higher $\pt$ region) where the fragmentation
of incoming partons is supposed to dominate hadron production.}

\section{Introduction}

\subsection{Strangeness Production at RHIC}
\label{subsec:strangeproduction}
In continuation of the long program of heavy ion collision studies
new measurements have been performed at RHIC, which successfully
switched from the first year (2000) energy of $\sqrt{s_{NN}}$=
$130~GeV$ to the maximum attainable $200~GeV$ (second run of 2001).
These energies are an order of magnitude higher than those achieved
in fixed target experiments at the SPS which makes it more likely
that a deconfined partonic phase is created. In order to unambiguously
identify such a state of matter, the initial conditions of $Au$+$Au$
collisions need to be characterised.

\noindent As strange quarks are created and not transported
from incoming nuclei, strangeness production is expected to be a good
estimator of the degree of equilibration of the produced fire-ball.
From their subsequent decays, strange particles can be clearly
identified  with a detector like STAR due to its large phase space
coverage complemented by a high efficiency tracking. 

\subsection{The STAR Experiment}
\label{subsec:stardetector}
The Solenoidal Tracker At RHIC (STAR) tracking system~\cite{star},
situated mainly at mid-rapidity, consists of a large cylindrical
Time Projection Chamber (TPC) assisted by an inner silicon
detector (SVT) and an outer RICH patch.
Charged particle tracks are bent by a longitudinal magnetic field
and extrapolated from the detection volume of the TPC toward the
SVT and the interaction point.
For each track, identification implies topological analyses,
specific ionization in the detector gas and/or Cherenkov imaging.  
Triggering is provided with the simultaneous use of a Central
Trigger Barrel and two Zero Degree hadronic Calorimeters~\cite{zdc} located
upstream and downstream of the detector along the beam axis.
This system was supplemented by two beam-beam counters during the
$p$+$p$ run. Detailed explanations related to topological analyses
in STAR are provided elsewhere~\cite{staratio,lambda}.
However, one should keep in mind that the strength of such a method
resides in the lack of high $\pt$ limitation rather than statistics.\\

\noindent
In this paper, the preliminary results shown correspond to both
$Au$+$Au$ and $p$+$p$ colliding systems. Emphasis will be placed on
strange baryon ratios, discussing: (i) Baryon transport using 
$\bbb$ ratios; (ii) the behaviour of the $\al/\la$ ratio as a 
function of $\pt$; (iii) mixed strange and multi-strange ratios
as a probe of chemical equilibrium.

\section{Initial conditions and baryonic number}
\label{sec:baryonRatio}

\begin{figure}[htb]
\begin{center}
\epsfig{figure=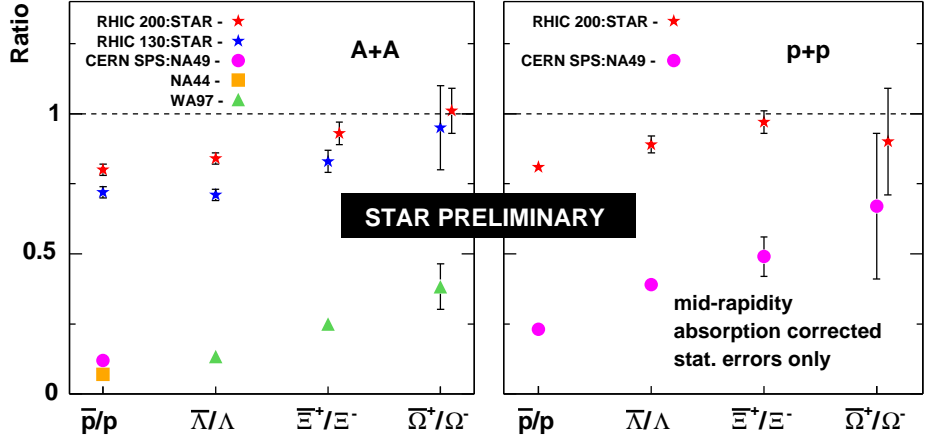,height=6cm}
\caption{Anti-baryon/baryon ($\bbb$) ratios as a function of
the strangeness content at RHIC and SPS energies.
Left panel corresponds to A+A most central collisions whereas
the right one corresponds to p+p.
These mid-rapidity ratios are corrected for absorption in the
detector material and the errors on the data are statistical only.
STAR ratios at $130~GeV$ are corrected for feed-down. 
\label{fig:baryonRatio}}
\end{center}
\end{figure}

Important information is encoded in the baryon number transport
mechanism, which corresponds to the very early stages of the collision
and affects the dynamic evolution of the matter created, more
specifically the thermal and chemical equilibration of the system. 
Net-baryon number has to be conserved from initial interactions
to the whole final rapidity interval. At mid-rapidity, the $\bbb$
ratio yields information on the corresponding net-baryon density, and 
one can simply compare the valence quark transport from beam
rapidities ($\sim 6$ units) to pair production.
As a function of the baryon strange quark content and beam energy,
a linear trend is clearly seen in Figure \ref{fig:baryonRatio}.

\noindent
Although the mid-rapidity region is not net-baryon free, the
ratios for $Au$+$Au$ at $200~GeV$ are even closer to unity than the
ones previously reported~\cite{staratio} at $130~GeV$ and of course
than those of the SPS experiments~\cite{spsratio} (left panel).
This implies that a still sizeable fraction of the baryon number
is transported from the incoming nucleus at beam rapidity to the
mid-rapidity region. 
A baryo-chemical potential of $\sim40~MeV$ ($\sim25~MeV$) for $130~GeV$ ($200~GeV$)
can be extracted with a widely used statistical formalism \cite{pbm}.
Hadronization is therefore thought to be located close to the region
where QCD calculations on the lattice are valid.    
It is interesting to attribute this behaviour to an energy dependence.
Indeed, in comparing $Au$+$Au$ (even from central to peripheral 
collisions~\cite{lambda}) and $p$+$p$ collisions at RHIC, both
at $200~GeV$, similar $\bbb$ values are obtained.\\

\section{Soft physics up to moderate transverse momentum}

\begin{figure}[htb]
\begin{center}
\epsfig{figure=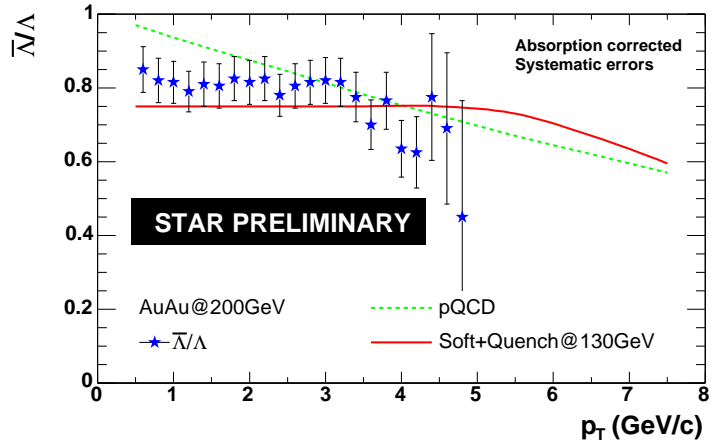,height=6cm}
\caption{STAR measurement of Anti-Lambda/Lambda as a function
of transverse momentum at $\sqrt{s_{NN}}$=$200~GeV$.
The ratio is consistent with a flat behavior up to moderate
$p_{T}\simeq4GeV/c$.
Calculations (see references in the text) for pQCD (solid) and
soft+quenching at $130~GeV$ (dashed) are shown for comparison.
\label{fig:ptRatio}}
\end{center}
\end{figure}

The integrated ratios presented earlier are very convenient since
no efficiency corrections are needed. The main assumption is that
the detection is charge symmetric and modulo absorption, the
efficiency for both charges should be the same.
However, the raw spectra are not usually flat as a function of $\pt$
(corrected is exponential-like and 95\% of the hadrons produced at RHIC
are below $2~GeV/c$) and the low $\pt$ part dominates the integrated
ratio value. Therefore it is legitimate to wonder how
these ratios behave as a function of $\pt$, especially because
the contribution of baryon transport and pair production should
affect strange and strange anti-baryons differently.
The example of $\la$ and $\al$ is interesting since similar 
production for both particles were previously reported~\cite{lambda}
at $130~GeV$.
Figure \ref{fig:ptRatio} shows STAR preliminary
results for the $\al/\la$ ratio at $200~GeV$. It is $\pt$-independent
up to a moderate value of $\sim 4~GeV/c$. Although the $\pt$ range
is smaller at $130~GeV$ due to statistics, similar results
are obtained for the $\axp/\xim$ ratio~\cite{staratio}. 

Recent investigations have been undertaken to explain baryon
dynamics using gluonic baryon junctions~\cite{vitev}. Corresponding
calculations but at $130~GeV$ are superimposed on the figure
where it appears the soft+quenching curve is closer to the STAR
preliminary data than pQCD inspired calculations.
Indeed, no constant decrease is seen for this ratio
whereas a turnover may occur around $3\sim4~GeV/c$.
Taking into account systematic errors, one needs to extend
the measurements at higher $\pt$ in order to draw a pertinent
conclusion. Similar baryon ratios (e.g. $\axp/\xim$) would 
certainly be useful for any confirmation of a transition between
a ``soft'' regime and the perturbative one.

\section{Highly equilibrated system at hadronization}

\begin{figure}[htb]
\begin{center}
\epsfig{figure=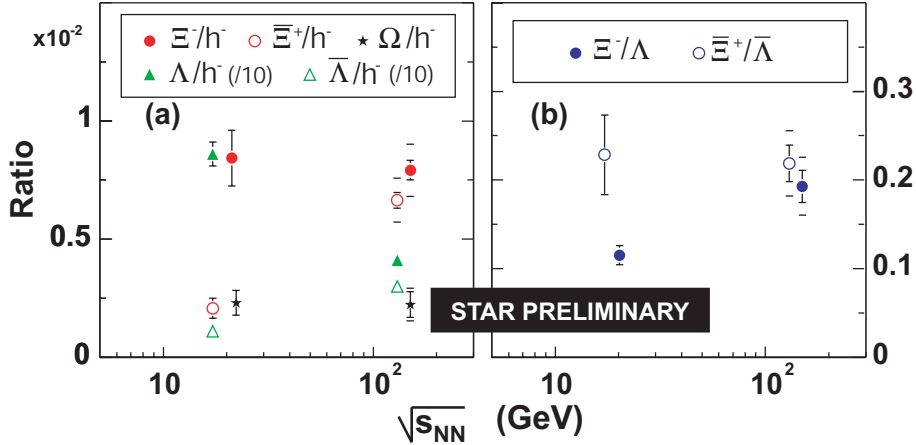,height=6cm}
\caption{ Mixed ratios as a function of the beam energy. (a) $\xim$,
$\axp$, $\Omega$=$\omm$+$\aop$, $\la$ and $\al$ to $h^{-}$; (b) $\xim/\la$
and $\axp/\al$. Statistical (solid lines) and systematic uncertainties
are added in quadrature (caps). Ratios involving $\xim$ and $\Omega$ are
slightly shifted along the x-axis for clarity.
\label{fig:mixedRatio}}
\end{center}
\end{figure}

Strange and multi-strange particle production as a function of
entropy is a very important tool for studying chemical equilibration.
It has been often shown how difficult it is to reproduce the
multi-strange baryon yields with microscopic models including
hadronic phases only ~\cite{hijing}.
Figure \ref{fig:mixedRatio} shows both strange and multi-strange
ratios as a function of beam energy. In the left panel (a),
the $\axp/h^{-}$ and $\al/h^{-}$ ratios for the most central
data increase from SPS energies to RHIC, whereas the
$\xim/h^{-}$ ratio stays constant and the $\la/h^{-}$ decreases.

As discussed in section \ref{sec:baryonRatio}, the reduction of
the net-baryon density between these two beam energies affect 
the $\la$ and the $\xim$ differently. Therefore, the behaviour of
these ratios with beam energy comes from both the increase of 
entropy and the decrease in net-baryon density. These effects
seem to cancel for the $\Omega=\Omega^{-}+\bar{\Omega}^{+}$.
However, it is interesting to note that the $\axp$/$\al$ ratio is
a constant from SPS to RHIC (see Figure \ref{fig:mixedRatio}(b)),
suggesting that the scale of the multi­strange enhancement is the
same for both singly and doubly strange baryons. 

\section{Conclusion}

A global picture emerges from the strange baryon production
in heavy ion collisions at RHIC. Net-baryon density
decreases with increasing collision energies but still differs from
zero. It means pair production is now dominant even if
baryon transport from beam subsists at mid-rapidity.
Strange particle yields are well reproduced by statistical
models, which strongly suggests a high degree of chemical
and thermal equilibrium at the hadronization stage.
The obtained parameters fit well into the region of temperatures
and baryo-chemical potentials where lattice QCD calculations
predict the phase transition. New measurements with better
statistics from STAR will help distinguishing the soft and
hard contributions in hadron production. Topological analyses
for strange baryons, which benefit from a potentially unlimited
$\pt$ coverage, will certainly contribute to this.   

\section*{Acknowledgments}
We wish to thank the RHIC Operations Group and the RHIC Computing Facility
at Brookhaven National Laboratory, and the National Energy Research 
Scientific Computing Center at Lawrence Berkeley National Laboratory
for their support. This work was supported by the Division of Nuclear 
Physics and the Division of High Energy Physics of the Office of Science of 
the U.S. Department of Energy, the United States National Science Foundation,
the Bundesministerium fuer Bildung und Forschung of Germany,
the Institut National de la Physique Nucl\'{e}aire et de la Physique 
des Particules of France, the United Kingdom Engineering and Physical 
Sciences Research Council, Fundacao de Amparo a Pesquisa do Estado de Sao 
Paulo, Brazil, the Russian Ministry of Science and Technology, the
Ministry of Education of China, the National Natural Science Foundation 
of China, and the Swiss National Science Foundation.


\section*{References}
\begin{thebibliography}{99}

\bibitem{star} K. H. Ackermann {\it et al}, \Journal{\NIM}{A499}{624}{2003}.

\bibitem{zdc}C. Adler {\it et al}, \Journal{\NIM}{A470}{488}{2001}.

\bibitem{staratio}J. Adams {\it et al}, submitted to \PLB, nucl-ex/0211024.

\bibitem{lambda}C. Adler {\it et al}, \Journal{\PRL}{89}{092301}{2002}.

\bibitem{spsratio} M. Kaneta {\it et al} (NA44), \Journal{J. Phys}{G23}{1865}{1997},\\ 
E. Andersen {\it et al} (WA97), \Journal{J. Phys}{G25}{171}{1999},\\
S.V Afanasiev {\it et al} (NA49), nucl-ex/0208014.

\bibitem{pbm}P. Braun-Munzinger {\it et al}, \Journal{\PLB}{518}{41}{2001}.

\bibitem{vitev}I. Vitev and M. Gyulassy, hep-ph0208108.

\bibitem{hijing}X.N. Wang and M. Guylassy, \Journal{\PRD}{44}{3501}{1991},\\
 S.E. Vance {\it et al} \Journal{\PLB}{448}{45}{1998}.

\end{thebibliography}
\end{document}